\documentclass[prl,twocolumn,showpacs,preprintnumbers,amsmath,amssymb]{revtex4}
\usepackage{graphicx}
\newcommand{\beq}{\begin{equation}}
\newcommand{\eeq}{\end{equation}}
\newcommand{\beqa}{\begin{eqnarray}}
\newcommand{\eeqa}{\end{eqnarray}}

\newcommand{\Sigs}{\Sigma_{\mathrm s} }
\newcommand{\Sigv}{\Sigma_{\mathrm v} }
\newcommand{\Sigo}{\Sigma_{\mathrm o} }

\newcommand{\kf}{k_{\mathrm F} }

\newcommand{\bfgamma}{\mbox{\boldmath$\gamma$\unboldmath}}
\newcommand{\bftau}{\mbox{\boldmath$\tau$\unboldmath}}

\newcommand{\veck}{\textbf{k}}

\newcommand{\vecq}{\textbf{q}}

\newcommand{\qabs}{|{\bf q}|}
\newcommand{\kabs}{|{\bf k}|}
\makeatletter
\newcommand{\fmslash}[2][0mu]{%
  \mathchoice
    {\fmsl@sh\displaystyle{#1}{#2}}%
    {\fmsl@sh\textstyle{#1}{#2}}%
    {\fmsl@sh\scriptstyle{#1}{#2}}%
    {\fmsl@sh\scriptscriptstyle{#1}{#2}}}
\newcommand{\fmsl@sh}[3]{%
  \m@th\ooalign{$\hfil#1\mkern#2/\hfil$\crcr$#1#3$}}
\makeatother

\begin{document}
\preprint{}
\title{Model independent study of the Dirac structure of the nucleon-nucleon interaction} 
\author{O. Plohl}
\author{C. Fuchs}
\author{E. N. E. van Dalen}
\affiliation{Institut
f$\ddot{\textrm{u}}$r Theoretische Physik, Universit$\ddot{\textrm{a}}$t
T$\ddot{\textrm{u}}$bingen,
Auf der Morgenstelle 14, D-72076 T$\ddot{\textrm{u}}$bingen, Germany}
\begin{abstract}
Relativistic and non-relativistic modern nucleon-nucleon potentials are 
mapped on a relativistic operator basis using projection techniques. 
This allows to compare the various potentials at the level of covariant 
amplitudes were a remarkable agreement is found. In nuclear matter 
large scalar and vector mean fields of several hundred MeV 
magnitude are generated at tree level. This is found to be a model 
independent feature of the nucleon-nucleon interaction. 
\end{abstract}
\maketitle

A fundamental question in nuclear physics is the role which 
relativity plays in nuclear systems. The ratio of the Fermi momentum 
over the nucleon mass is about $k_{\rm F}/M \simeq 0.25$ and 
nucleons move thus with maximally about 1/4 of the velocity of light which  
implies only moderate corrections from relativistic kinematics. 
However, there exists a fundamental difference between relativistic 
and non-relativistic dynamics: a genuine feature of relativistic 
nuclear dynamics is the appearance of large scalar and vector 
mean fields, each of the magnitude of several hundred MeV. The scalar 
field $\Sigma_S$ is attractive and the vector field  $\Sigma_\mu $ is 
repulsive. In relativistic mean field (RMF) theory, both, sign 
and size of the fields are enforced by the nuclear 
saturation mechanism \cite{sw86}.  

At nuclear saturation density $\rho_0\simeq 0.16~{\rm fm}^{-3}$ 
the empirical fields deduced from RMF fits to finite nuclei are 
of the order of  $\Sigs\simeq -350$ MeV and 
$\Sigo \simeq +300$ MeV \cite{rmf} ($\Sigo$ is the time-like 
component of $\Sigma_\mu$ ). 
The single particle potential in which the nucleons move 
originates from the cancellation of the two contributions 
$U_{\rm s.p.} \simeq \Sigo + \Sigs \simeq -50$ MeV which 
makes it difficult to observe relativistic effects in 
nuclear systems. There exist, however, several 
features in nuclear structure which can naturally be explained 
within Dirac phenomenology while models based on non-relativistic dynamics 
have difficulties. Best established 
is the large {\it spin-orbit splitting} in 
finite nuclei. Also the so-called {\it pseudo-spin symmetry}, observed 
more than thirty years ago in single-particle 
levels of spherical nuclei, can naturally be understood within RMF 
theory as a consequence of the coupling to the lower components of 
the Dirac equation \cite{ginocchio97}.

A connection to Quantum-Chromo-Dynamics (QCD) 
as the fundamental theory of strong interactions 
is established by QCD sum rules \cite{cohen91,drukarev91}. The change  
of the chiral condensates 
$\langle {\bar q}q \rangle, \langle q^\dagger q \rangle $ in matter leads to 
attractive scalar and repulsive vector self-energies which 
are astonishingly close to the empirical values 
derived from RMF fits to the nuclear chart. 
Also relativistic many-body calculations 
\cite{terhaar87a,gross99,dalen04}  yield 
scalar/vector fields of the same sign and magnitude 
as obtained from RMF theory or, alternatively, from QCD sum rules. 
Moreover, Dirac-Brueckner-Hartree-Fock (DBHF) 
calculations  \cite{gross99} agree even 
quantitatively surprisingly 
well with the  QCD motivated approach of Ref. 
\cite{finelli} where chiral fluctuations from 
pion-nucleon dynamics were considered 
on top of the chiral condensates. 

These facts suggest that preconditions for the 
existence of large fields in matter or, 
alternatively, the density dependence of the QCD condensates, must already 
be inherent in the vacuum nucleon-nucleon (NN) interaction. 

Relativistic one-boson-exchange potentials (OBEP), e.g. Bonn (A,B,C), 
\cite{bonn} provide fits with fair precision 
to NN scattering data. High precision fits 
(with $\chi^2/{\rm datum}\simeq 1.01$) are provided  
by OBE type  potentials such as the relativistic CD-Bonn \cite{cdbonn} 
or the Nijmegen potentials \cite{nijmegen} where fits are 
performed separately in each partial wave. In contrast to e.g. 
the spectator model of \cite{gross92} the 'standard' OBE 
potentials are based on the no-sea approximation which excludes 
explicit excitations of anti-nucleons but absorbs such contributions 
into the model parameters, in particular into a large 
$\omega$ coupling. The present investigations are 
restricted to the standard-type models.

The long-range part of the interaction is generally 
mediated by the one-pion-exchange (OPE) while the scalar intermediate range
attraction is mainly due to correlated two-pion-exchange. The short-range
part, i.e. the hard core, is dominated by the heavy vector meson exchange, 
i.e. the isoscalar $\omega$ and the isovector $\rho$ meson. 
Widely used in nuclear physics are, however, also 
high precision non-relativistic empirical potentials such as the 
Argonne potential AV$_{18}$ \cite{av18}  . 
A systematic connection to QCD is established by 
chiral effective field theory (EFT). Up to now the two-nucleon system 
has been considered at next-to-next-to-next-to-leading order (N$^3$LO) in 
chiral perturbation theory \cite{entem02,epelbaum05}. 
In such approaches 
the NN potential consists of one-, two- and three-pion exchanges and 
regularizing contact interactions which 
account for the short range correlations. 
The advantage of EFT compared to the more phenomenological 
OBE potentials is the systematic expansion 
in terms of chiral power counting. 

A better understanding of the common features of the 
various approaches is essential in order to arrive at a more model independent 
understanding of the nucleon-nucleon interaction, in particular since 
all well established interactions fit NN scattering data with 
approximately the same precision. Substantial progress was recently 
achieved by the construction of a universal low energy NN 
potential $V_{\rm low~k}$ based on renormalization group 
techniques \cite{lowk03}. However, like the EFT potentials 
 $V_{\rm low~k}$ is not covariantly formulated and has therefore 
not been used in relativistic nuclear structure calculations.  

The present work applies projection techniques to map the 
various potentials on the operator basis of relativistic field theory 
which is given by the Clifford algebra in Dirac space. 
This allows to identify the different Lorentz 
components of the interaction and to calculate the relativistic 
self-energy operator in matter. The 
philosophy behind this approach is based on the fact, that any NN interaction, 
whether relativistic or not, is based on a common  
spin-isospin operator structure which invokes certain scales:  
long-range spin-isospin dependent forces, essentially given  by the 
one-pion exchange, short- and intermediate-range spin-independent 
interactions, short-range isoscalar spin-orbit interactions and 
quadratic isovector spin-orbit interactions. 

In a covariant formulation the  spin-isospin structure of the 
interaction is constrained by the symmetries of the Lorentz group. 
E.g. the  Born scattering matrix of covariant OBE potentials 
is given by the sum over the corresponding scalar, pseudoscalar and 
vector mesons $\alpha$
\begin{equation}
{\hat V} (q^\prime , q)
=\sum_{\alpha=s,ps,v} {\cal F}^{2}_{\alpha}(q^\prime , q)~ 
\Gamma_{\alpha}^{(2)} ~
D_{\alpha}(q^\prime - q) ~\Gamma_{\alpha}^{(1)} ~~,
\label{vobe1}
\end{equation}
where $q_\mu$ and $q_{\mu}^\prime$ are the c.m. momenta 
of the incoming and outgoing 
nucleons. The meson propagator is denoted by $D_\alpha$, ${\cal F}_{\alpha}$ are form factors applied to the vertices $\Gamma_{\alpha}$.
Isospin factors ($\bftau_1\!\cdot\!\bftau_2$) are suppressed in 
(\ref{vobe1}). 
 The Dirac structure of the potential is contained in the 
meson-nucleon vertices 
\begin{eqnarray}
\Gamma_{s}=g_{s}\large{\bf{1}},~   
  \Gamma_{ps}=g_{ps}
\frac{\fmslash{q}^\prime - \fmslash{q}}{2M} i\large{\gamma_5},~
 \Gamma_{v}=g_v\large{\gamma^{\mu}}
+ \frac{f_v i\sigma^{\mu\nu}  }{2M}~.
\label{vertex1}
\end{eqnarray}
For the pseudoscalar mesons $\pi$ and $\eta$ a pseudovector coupling 
is used and the $\omega$ meson has no tensor coupling, i.e. 
$f_{v}^{(\omega)} =0$. The potential $V ({\bf q}',{\bf q})$, 
i.e. the OBE Feynman amplitudes are obtained by 
sandwiching ${\hat V}$ between the incoming and outgoing Dirac spinors. 

From a low energy expansion of the OBE amplitudes one obtains 
the representation of non-covariant potentials 
\begin{equation}
V({\bf q'},{\bf q})= \sum_{\alpha} [V_{\alpha}+V_{\alpha}'\,\bftau_1\!\cdot\!\bftau_2]\,\,O_{\alpha} 
\label{argonne2}
\end{equation}
where the most important operators (assuming identicle particles) are 
\begin{eqnarray}
   O_{1}&=&1,   \quad   
   O_{2}=\mbox{\boldmath $\sigma$}_{1}\!\cdot\!
         \mbox{\boldmath $\sigma$}_{2}, \quad 
\nonumber \\
   O_{3}&=&(\mbox{\boldmath $\sigma$}_{1}\!\cdot\!{\bf k})
         (\mbox{\boldmath $\sigma$}_{2}\!\cdot\!{\bf k})
       - {\textstyle\frac{1}{3}}{\bf k}^{2}
         (\mbox{\boldmath $\sigma$}_{1}\!\cdot\!
          \mbox{\boldmath $\sigma$}_{2}),   
\nonumber \\
   O_{4}&=&{\textstyle\frac{i}{2}}
        (\mbox{\boldmath $\sigma$}_{1}+\mbox{\boldmath $\sigma$}_{2})
        \cdot{\bf n}, \quad 
   O_{5}=(\mbox{\boldmath $\sigma$}_{1}\!\cdot\!{\bf n})
         (\mbox{\boldmath $\sigma$}_{2}\!\cdot\!{\bf n}), 
\label{Pmom}
\end{eqnarray}
with ${\bf k}={\bf q'}-{\bf q}$, ${\bf P}=\frac{1}{2}({\bf q'}+{\bf q})$ 
and ${\bf n}={\bf P}\times {\bf k}$.

The modern non-covariant Nijm I and II potentials are 
constructed from approximate OBE amplitudes and thus 
based on the operator structure (\ref{Pmom}). However, 
since they are separately fitted in each partial wave 
they are often referred to as phenomenological models.
The Argonne potential AV$_{18}$ \cite{av18} is purely 
phenomenological in the sense that the OBE picture is 
released. Based on the symmetries of (\ref{Pmom}), 
 the intermediate and short range part is parameterised by 
phenomenological functions $V_{\alpha}$ and only 
one-pion-exchange is contained explicitly.
The EFT potentials, containing one- and multi-pion 
exchange explicitely, are even more rigorous since most part 
of the nuclear repulsion is carried by regularizing counter terms. 
Such an approach is justified by the fact that heavy meson ($\rho,\omega$) 
exchange cannot be resolved up to the momentum scale of about 400 MeV 
which is constrained by NN scattering data. We apply the 
EFT Idaho potential \cite{entem02} which fits 
NN scattering data with similar quality as  AV$_{18}$ or the Nijmegen 
potentials. The same philosophy 
is behind $V_{\rm low~k}$ which can be viewed as the condensation 
of the various formulations to a model independent result. 

Although based on the same operator structure as OBEPs, 
it is not straightforward to fix the Lorentz character of 
the various pieces of the non-covariant 
interactions AV$_{18}$, EFT (Idaho) and $V_{\rm low~k}$. 

However, any two-body amplitude can be represented covariantly by 
Dirac operators and Lorentz invariant amplitudes \cite{tjon85}. 
A relativistic treatment invokes 
automatically the excitation of anti-nucleons. However, NN scattering, 
in both, relativistic and  non-relativistic 
approaches is restricted to the positive 
energy sector and neglects the coupling to anti-nucleons. As a consequence 
one has to work in a subspace of the full Dirac space where 
on-shell two-body matrix elements
can be expanded into five Lorentz invariants. A possible choice of a set of five
linearly independent covariant operators are 
the scalar, vector, tensor, axial-vector and 
pseudo-scalar Fermi covariants $\Gamma_m =\{{\rm S,V,T,P,A}\}$ with 
$ {\rm S}=\large{\bf{1}}\otimes\large{\bf{1}}, ~ {\rm V}=\gamma^{\mu}\otimes\gamma_{\mu},~ 
  {\rm T}=\sigma^{\mu\nu}\otimes \sigma_{\mu\nu},~
{\rm P}=\gamma_5\otimes\gamma_5, ~
{\rm A}=\gamma_5\gamma^{\mu}\otimes \gamma_5\gamma_{\mu}$. Working with physical,
i.e. antisymmetrized matrix elements $V^A$, one has to keep in mind that 
direct and exchange covariants ${\tilde \Gamma}_m$ (where 
Dirac indices of particles 1 and 2 are interchanged) are coupled 
by a Fierz transformation. Since the low momentum part of the 
NN interaction is totally dominated by the pseudovector 
one-pion exchange (OPE) it is preferable to choose an operator 
basis where the pseudovector exchange is completely separated from the 
remaining  operator structure. This can be achieved by the 
following set of covariants originally proposed 
by Tjon and Wallace \cite{tjon85} 
\beq
\Gamma_m = \{ {\rm S},-{\rm {\tilde S}},({\rm A}-  {\rm {\tilde A}}),{\rm PV} ,
-{\rm {\widetilde {PV}}}\}~.
\label{cov3}
\eeq
PV and ${\rm {\widetilde {PV}}}$ are the direct and exchange 
pseudovector covariants, analogous to the pseudoscalar covariant P, 
however, with $\gamma_5$ replaced by 
$(\fmslash{q}^\prime - \fmslash{q})/2M \gamma_5$. Thus the on-shell 
($|{\bf q}|=|{\bf q}^\prime|$) scattering matrix is given by 
\begin{equation}
{\hat V}^I ({\bf q}^\prime ,{\bf q})
=\sum_{m} g_{m}^I(|{\bf q}|,\theta)~ \Gamma_m~~,
\label{vcov}
\end{equation}
where $\theta$ is the c.m. scattering angle and $I=0,1$ the 
isospin channel. For the Hartree-Fock self-energy 
it is sufficient to consider $\theta=0$  
when antisymmetrized matrix elements are used since 
 $\theta=\pi$ contains then only redundant information. 
The transformation of the Born 
amplitudes from an angular-momentum basis onto 
the covariant basis (\ref{vcov}) is now standard and 
runs over the following steps:  
$ |LSJ\rangle \rightarrow$ {\it partial wave helicity states 
$ \rightarrow$ plane wave helicity states $\rightarrow$ 
covariant basis}. The first two transformation can e.g. be found in Refs. 
\cite{machleidt89}. The last step has to be performed numerically 
by matrix inversion  \cite{horowitz87,gross99}.  

\begin{figure}[htb]
\includegraphics[width=0.5\textwidth] {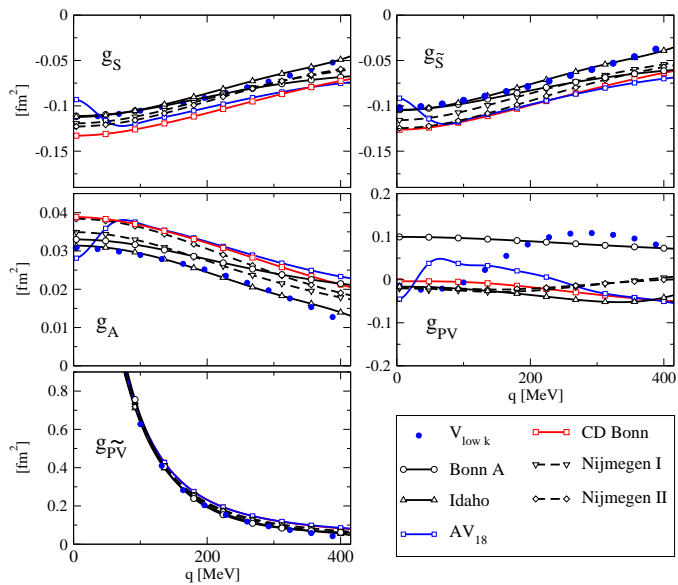}
\caption {Isospin-averaged Lorentz invariant
  amplitudes $g_m(|{\bf q}|,0)$ for the 
different NN potentials after projection on the Dirac 
operator structure. The pseudovector representation of the 
relativistic operator basis is used. 
\label{F_pv_d_Itotal} }
\end{figure}
Fig. \ref{F_pv_d_Itotal} shows the resulting isospin-averaged amplitudes 
 $g_m = \frac{1}{2}(g_{m}^{I=0}+ 3g_{m}^{I=1})$ at $\theta=0$ 
for the various potentials up to 400 MeV. 
The four amplitudes  $ g_{\rm S},~g_{\tilde {\rm S}},~ g_{\rm A}$     
and  $ g_{\widetilde {\rm PV}}$ are very close for the OBEPs 
 Bonn A, CD-Bonn, Nijmegen I and II, and the non-relativistic 
AV$_{18}$, Idaho and  $V_{\rm low~k}$. The direct 
pseudovector amplitude  $g_{\rm PV}$ falls somewhat out of systematics. 
This amplitude is, however, of minor 
importance since it does not contribute to the Hartree-Fock 
self-energy (\ref{sighf}) and to the single particle 
potential. The dominance of the OPE at low $|{\bf q}|$ is reflected in the pseudovector 
exchange amplitude $ g_{\widetilde {\rm PV}}$ which is at small $|{\bf q}|$ 
almost two orders of magnitude larger than the other amplitudes. 
For the covariant OBEPs Bonn and CD-Bonn and the OBE-like 
Nijmegen potentials a coincicdence at the level of covariant 
amplitudes is not completely unexpected. However, it is remarkable 
that this agreement transfers to the EFT Idaho potential and 
also to  $V_{\rm low~k}$ which are of completely different character 
and theoretical background.   

With the covariant amplitudes at hand, one is now able to determine 
the relativistic mean field in nuclear matter by calculating the relativistic self-energy $\Sigma$ in Hartree-Fock 
approximation at {\it tree level}. We are thereby not aiming for a realistic
description of nuclear matter saturation properties which would require 
a self-consistent scheme. Moreover,  short-range correlations require 
to base such calculations on the in-medium T-matrix rather than the 
bare potential $V$ \cite{muether00}. This leads to the relativistic 
Dirac-Brueckner-Hartree-Fock scheme which has been proven to 
describe nuclear saturation with quantitatively satisfying accuracy 
\cite{terhaar87a,gross99,dalen04}.

 The self-energy for the nucleon $k$ follows from the interaction matrix $V$ by  integrating over the occupied states $q$ in the Fermi sea
\beq
\Sigma_{\alpha\beta}(k,k_F)=-i\int {d^4q\over {(2\pi)^4}}~
G^D_{\tau\sigma}(q)~V^A(k,q)_{\alpha\sigma;\beta\tau}~.
\label{sighf}
\eeq
The Dirac propagator 
$G^D(q)=[\fmslash{q}+M]2\pi
 i\delta(q^2-M^2)\Theta(q_0)\Theta(k_F-\qabs)$ 
describes the on-shell propagation
of a nucleon with momentum ${\bf q}$ and energy $E_{\bf q}=\sqrt{{\bf
 q}^2+M^2}$ inside the Fermi sea.  
In isospin saturated  nuclear matter
the self-energy consists of a scalar $\Sigs$, a time-like 
vector $\Sigo$ and a spatial vector part $\Sigv$ 
\beqa
\Sigma(k,\kf)= \Sigs (k,\kf) -\gamma_0 \, \Sigo (k,\kf) + 
\bfgamma  \cdot \textbf{k} \,\Sigv (k,\kf),
\label{subsec:SM;eq:self1}
\eeqa
which are given by \cite{gross99}
\beqa
\Sigs  & = & \frac{1}{4} \int^{k_F} \!\!\!\!\!\frac{d^3\vecq}{(2
  \pi)^3}  \frac{M}{E_{\bf q}}  \left[ 4g_{\rm S} -
  g_{\tilde {\rm S}}+ 4 g_{\rm A} -\frac {(k^{\mu}-q^{\mu})^2} {4M^2}
  g_{\widetilde {\rm PV}} \right]~, \nonumber\\
\Sigo   & = &  \frac{1}{4} \int^{k_F}\!\!\!\!\! \frac{d^3\vecq}{(2 \pi)^3}    \left[ g_{\tilde {\rm S}}- 2 g_{\rm A}+ \frac{E_{\bf k}}{E_{\bf q}} \frac {(k^{\mu}-q^{\mu})^2} {4M^2} g_{\widetilde {\rm PV}}\right]
\label{sigHF}\\
\Sigv  & = & \frac{1}{4} \int^{k_F}\!\!\!\!\! \frac{d^3\vecq}{(2
  \pi)^3} \frac{\veck\cdot\vecq}{\kabs^2E_{\bf    q}}  
\left[ g_{\tilde {\rm S}}-2 g_{\rm A} + \frac{k_z}{q_z} \frac{(k^{\mu}-q^{\mu})^2} {4M^2}
  g_{\widetilde {\rm PV}} \right]
\nonumber 
\eeqa
\begin{figure*}[htb]
\includegraphics[width=0.85\textwidth] {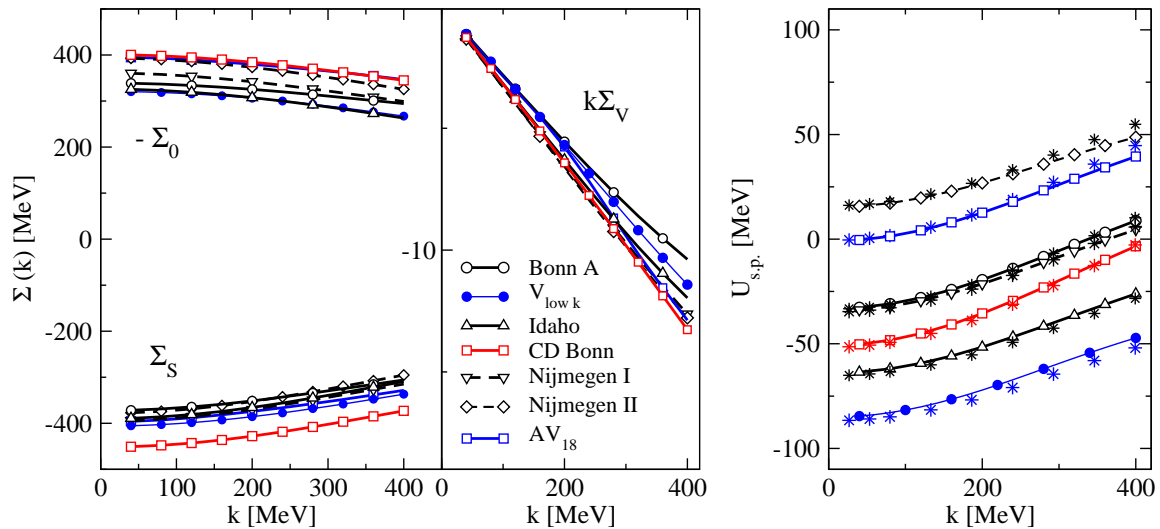}
\caption {Tree level results for nuclear matter at $\kf =1.35~{\rm fm}^{-1}$ 
for the various potentials: scalar and vector 
self-energy components (left), spatial 
vector self-energy component (middle) and single particle 
potential $U_{\rm s.p.}$ (right). Stars in the right panel denote 
a non-relativistic calculation of  $U_{\rm s.p.}$. 
\label{sigma_fig} }
\end{figure*}
Fig. \ref{sigma_fig} shows the tree level self-energy 
components obtained with the various potentials 
in nuclear matter at saturation density with a Fermi momentum 
$\kf =1.35~{\rm fm}^{-1}$. All potentials 
yield scalar/vector fields $\Sigs$ and   $\Sigo$ of comparable 
strength: a large and attractive scalar field  $\Sigs \simeq -(450\div 400)$
MeV and a repulsive vector field of $\Sigo \simeq +(350\div 400)$ MeV. 
The spatial components ${\bf k}\Sigv$ agree as well. 
As known from self-consistent DBHF calculations \cite{terhaar87a,gross99}, 
the spatial vector self-energy is a moderate correction to the large 
scalar and time-like vector components $\Sigs$ and $\Sigo$. This is 
already the case  at tree level. The right part of 
Fig. \ref{sigma_fig} shows the single particle potential 
$U_{\rm s.p.} (k,\kf) = M\Sigs/E_{\bf k}  - \Sigo + \Sigv {\bf k}^2/E_{\bf
  k}$, defined as the expectation value of the self-energy.  
It reflects the well known fact that 
various two-body potentials are rather different although they 
are phase-shift equivalent, i.e. they describe NN scattering data 
with about the same accuracy  
when iterated in the Lippmann-Schwinger equation \cite{muether00}. The 
differences at tree level are mainly due to differences in the 
short-range part of the interaction \cite{lowk03}. 
Fig. \ref{sigma_fig}  
includes also the results from a 'non-relativistic' calculation of 
$ U_{\rm s.p.}$  where the partial wave  amplitudes are 
directly summed up. The non-relativistic and the relativistic 
calculations show an excellent agreement which demonstrates 
the accuracy of the applied projection techniques. 

We presented a model independent study of the Dirac structure of the 
nucleon-nucleon interaction. The potentials were projected on a relativistic 
operator basis in Dirac space using standard projection 
techniques which transform from a partial wave basis, 
where the potentials are originally given, to the basis 
of covariant amplitudes. The different approaches can now 
be compared at the level of these  covariant amplitudes and, moreover, 
this allows to calculate the relativistic self-energy operator in nuclear 
matter. For both, the  covariant amplitudes and the tree-level Hartree-Fock 
self-energy, we observe a remarkable agreement 
between the OBEPs (Bonn, CD-Bonn, Nijmegen),   
the phenomenological AV$_{18}$ potential and the EFT based 
Idaho and $V_{\rm low~k}$ potentials. 
The structure of the interaction enforces 
the existence of large scalar and vector fields as a model 
independent fact. The scale of these fields is set at the tree level. Although 
essential for nuclear binding and saturation, higher order correlations, 
i.e. Brueckner ladder correlations, change the size of the fields by 
less than 20\% \cite{gross99}. The magnitude of the fields is similar to 
that predicted by relativistic 
mean field phenomenology, relativistic many body correlations and also 
by QCD sum rules. We conclude that the appearance of large scalar and 
vector fields in matter is a general and model independent 
consequence of  the vacuum structure of the NN interaction. Relativistic
dynamics is therefore essential for nuclear systems. 

\acknowledgments
We like to thank D. Entem for providing the Idaho N3LO program 
package and H. M\"uther for providing his Hartree-Fock 
code.



\begin{thebibliography}{99}

\bibitem{sw86} 
        B.D. Serot, J.D. Walecka,  
        Adv. Nucl. Phys. {\bf 16}, 1 (1986).
 
\bibitem{rmf} 
P. Ring, Prog. Part. Nucl. Phys. {\bf 73},  193 (1996).


\bibitem{ginocchio97}
J.N. Ginocchio, Phys. Rev. Lett. {\bf 78}, 436 (1997).

\bibitem{cohen91}
 T.D. Cohen, R.J. Furnstahl, D.K. Griegel, 
Phys. Rev. Lett. {\bf 67}, 961  (1991); Phys. Rev. C {\bf 45}, 1881 (1992).

\bibitem{drukarev91}
E.G. Drukarev, E.M. Levin, Prog. Part. Nucl. Phys. {\bf 27}, 77 (1991).

\bibitem{terhaar87a}
B. ter Haar and R. Malfliet,
Phys. Rep. {\bf 149}, 207 (1987).

\bibitem{gross99}
T. Gross-Boelting, C. Fuchs, and Amand Faessler,
Nucl. Phys. {\bf A648}, 105 (1999).

\bibitem{dalen04}
E. van Dalen, C. Fuchs, A. Faessler, Nucl. Phys. {\bf A744} 227 (2004); 
Phys. Rev. Lett. {\bf 95}, 022302  (2005).

\bibitem{finelli} 
P. Finelli, N. Kaiser, D. Vretenar, W. Weise,  
Nucl. Phys. {\bf A735}, 449 (2004). 

\bibitem{bonn}
R. Machleidt, K. Holinde, Ch. Elster, Phys. Rep.  {\bf 149}, 1 (1987).

\bibitem{cdbonn}
R. Machleidt, Phys. Rev. C {\bf 63}, 024001 (2001).



\bibitem{nijmegen}
V.G.J. Stoks, R.A.M. Klomp, M.C.M. Rentmeester, J.J. de Swart 
Phys. Rev. C {\bf 48}, 792 (1993).

\bibitem{gross92}
F. Gross, J.W. Van Orden, K. Holinde
Phys. Rev.C {\bf 45}, 2094 (1992).




\bibitem{av18}
R.B. Wiringa, V.G.J. Stoks, R. Schiavilla, 
Phys. Rev. C {\bf 51}, 38 (1995).

\bibitem{entem02} 
D.R. Entem, R. Machleidt, Phys. Rev. C {\bf 66}, 014002 (2002) 014002; 
Phys. Rev. C {\bf 68}, 041001 (2003).

\bibitem{epelbaum05}
E. Epelbaum, W. Gl\"ockle, U.-G. Meissner, 
Nucl. Phys. {\bf A747}, 362 (2005). 

\bibitem{lowk03}
S.K. Bogner, T.T.S. Kuo, A. Schwenk, Phys. Rep. {\bf 386}, 1 (2003). 
        


\bibitem{tjon85}
J.A. Tjon, S.J. Wallace,
Phys. Rev. C {\bf 32}, 267 (1985); Phys. Rev. C {\bf 32}, 1667 (1985).

\bibitem{machleidt89}
R. Machleidt, Adv. Nucl. Phys. {\bf 19}, 189 (1989).

\bibitem{horowitz87}
C.J. Horowitz, B.D. Serot,
Nucl. Phys. {\bf A464}, 613 (1987).


\bibitem{muether00} 
H. M\"uther, A. Polls, Prog. Part. Nucl. Phys. {\bf 45}, 243 (2000). 

\end{thebibliography}
\end{document}